\documentclass[aps,amsfonts,prd,preprint,nofootinbib,tightenlines]{revtex4}

\usepackage{epsf}

\def\be{\begin{equation}}
\def\ee{\end{equation}}
\def\bea{\begin{eqnarray}}
\def\eea{\end{eqnarray}}

\def\Im{\mathop{\rm Im}}

\def\sech{\mathop{\rm sech}\nolimits}

\begin{document}

\title{Negative Energy Densities in Quantum Field Theory With a Background
Potential}

\author{Noah\ Graham}
\email{ngraham@middlebury.edu}
\affiliation{Department of Physics,
Middlebury College, Middlebury, VT  05753}

\author{Ken D.\ Olum}
\email{kdo@cosmos.phy.tufts.edu}
\affiliation{Department of Physics and Astronomy,
Tufts University, Medford, MA  02155}

\preprint{hep-th/0211244}

\begin{abstract}

We present a general procedure for calculating one-loop ``Casimir'' energy
densities for a scalar field coupled to a fixed potential in renormalized
quantum field theory.  We implement direct subtraction of counterterms
computed precisely in dimensional regularization with a definite
renormalization scheme.  Our procedure allows us to test quantum field
theory energy conditions in the presence of background potentials
spherically symmetric in some dimensions and independent of others.
We explicitly calculate the energy density for several examples.  For a
square barrier, we find that the energy is negative and divergent
outside the barrier, but there is a compensating divergent positive
contribution near the barrier on the inside.  We also carry out
calculations with exactly solvable $\sech^2$ potentials, which arise
in the study of solitons and domain walls.
\\ {\rm PACS 03.65.Nk 11.10.Gh 11.27.+d 11.55.Hx}
\end{abstract}

\maketitle

\section{Introduction}

The weak energy condition of general relativity, the requirement that there
be no negative energy densities, is sufficient to prevent the appearance of
exotic features such as compactly generated closed timelike curves
\cite{Hawking} and superluminal travel \cite{superluminal}.  Quantum field
theory appears to violate this condition, however.  One example is the
standard Casimir system of parallel plates, for which there is a negative
energy density between the plates.  However, this is an idealized system,
which assumes a perfect conductor with an infinitely sharp and flat edge. 
A real material will have a rough surface at the atomic scale and will also
appear transparent to very high energy modes.  Since the Casimir energy is
a sum over all energies, it will always include modes for which these
effects are relevant, so this idealization could affect the value of the
sum.  In addition, since the boundary condition is imposed externally,
there is no measure of the energy that would be required to maintain
it.\footnote{Recently, Helfer and Lang \cite{HelferLang} showed that a
frequency-independent dielectric would not be expected to give negative
energy densities, but Sopova \cite{Sopova} showed that negative energy
densities can be achieved in a Casimir system with Drude-model plates, as
long as the spacing is very large compared to the plasma wavelength.}

Although there are other ways to produce negative energy densities,
for example a superposition of states with zero and two photons, these
cases are constrained by averaged energy conditions, which require the
energy to be positive when averaged along an entire geodesic.  They are
also constrained by quantum inequalities \cite{FordRoman},
which limit the total negative energy that can exist when averaging
over a certain period of time.  Thus it is important to understand
problems of the Casimir type if we want to know whether quantum field
theory protects general relativity against negative energies.

In this paper, we reconsider the question of the energy density in such
systems.  To avoid the subtleties associated with the Casimir problem, we
consider a quantum field in the presence of a background potential (i.e., a
field with a mass that depends on position).  In such an approach, one can
choose a potential that depends only on one spatial dimension, and simulate
the parallel plates in the Casimir problem.  By generalizing the approach
of \cite{domainwall,method,method2} to local densities, we can precisely cancel
the divergences in the calculation in a definite renormalization scheme.

Ref.\ \cite{SUNYSB} considered similar problems for the special case of
reflectionless potentials, such as the potentials for the supersymmetric
kink and sine-Gordon models.  Here we present a general approach suitable
for numerical computation, in addition to analytic calculations in exactly
solvable models \cite{letter}.  These techniques are also useful for the
study of Casimir forces and stresses \cite{MITgroup}.

\section{A simple model}

To illustrate our method, we will first consider a simple model.  We
take a real, massless scalar field in 2+1 dimensions in the
background of a repulsive potential $V$ that depends on one spatial dimension
but not the other.

We start with the Hamiltonian density
\be
{\cal H} ={1\over 2}\left[\dot\phi^2+(\nabla\phi)^2+ V\phi^2\right]
\ee
and expand the field $\phi$ in terms of small oscillations, giving
\be
\phi(x, y, t) =\int_{-\infty}^\infty{dp\over \sqrt{2\pi}}
\int_0^\infty{dk\over\sqrt{2\pi\omega (p)}}\sum_{\chi=+, -}
\left(\psi_k^\chi(x) e^{ipy} e^{-i\omega (p) t}a_{k,p}^\chi
+\psi_k^\chi(x)^\ast e^{-ipy} e^{i\omega (p) t}{a^\chi_{k,p}}^\dag\right) \,.
\ee
where $\omega(p) = \sqrt{k^2+p^2}$.  The $\psi_k^\chi (x)$ are normal
mode wave functions, which can be taken to be real.  The sum is over the
symmetric mode, $\psi^+_k(x)$, and the antisymmetric mode, $\psi^-_k(x)$. 
They satisfy
\be\label{eqn:eom}
-{\psi^\chi_k}''(x) +V(x)\psi_k^\chi(x) = k^2\psi_k^\chi (x)
\ee
where prime denotes differentiation with respect to $x$, with the
normalization
\be
\int_{-\infty}^\infty dx \,
\psi_k^+(x) \psi_{k'}^-(x) = 0 \qquad \hbox{and} \qquad
\sum_{\chi=+,-} \int_{-\infty}^\infty dx
\psi_k^\chi (x) \psi_{k'}^{\chi}(x) = 2\pi\delta (k-k') \,,
\ee
which gives ${\psi_k^+}^{(0)}(x) =\cos kx$ and ${\psi_k^-}^{(0)}(x)
=\sin kx$ as the solutions in the free case, $V(x) = 0$.

The energy density is then
\be
\langle{\cal H}\rangle =
\int_{-\infty}^\infty{dp\over 2\pi}
\int_0^\infty{dk\over 2\pi\omega (p)}\sum_{\chi=+, -} {1\over 2}
\left[(\omega (p)^2 + p^2 + 
V(x))\psi^\chi_k(x)^2+{\psi^\chi_k}'(x)^2\right] \,.
\ee
We write
\be
\psi'(x)^2 ={1\over 2}{d^2\over dx^2} \psi(x)^2-\psi''(x)\psi(x)
\ee
and then use Eq.\ (\ref{eqn:eom}) to obtain
\be
\langle{\cal H}\rangle =
\int_{-\infty}^\infty{dp\over 2\pi}
\int_0^\infty{dk\over 2\pi\omega (p)}\sum_{\chi=+, -}
\left[\omega (p)^2 \psi_k^\chi(x)^2 +
{1\over 4}{d^2\over dx^2}\left(\psi_k^\chi(x)^2\right)\right] \,.
\ee
The integral is highly divergent, but by using dimensional
regularization and introducing counterterms into the Hamiltonian, as
discussed in Sec. \ref{sec:method}, we can render the integral finite.
We then integrate out the transverse modes, giving
\bea
\langle{\cal H}\rangle_{\text{ren}} =
-\frac{\Gamma\left(-\frac{n+1}{2}\right)}{2(4\pi)^{\frac{n+1}{2}}}
\int_0^\infty \frac{dk}{\pi} \, \sum_{\chi=+, -}
\bigg\{&&\omega^{n+1}\left[\psi_k^\chi(x)^2-\left(1+{V(x)\over 2k^2}\right)
{\psi_k^\chi}^{(0)}(x)^2\right]\cr
&& +{n+1\over 4}\omega^{n-1}{d^2\over
dx^2}\left(\psi_k^\chi(x)^2\right)\bigg\} \,,
\eea
where $n$ is the number of transverse dimensions, later to be set to
1. The Hamiltonian has been renormalized by subtracting a constant term for
the vacuum energy and a term proportional to the potential,
which is sufficient to render it finite in 2+1 dimensions.

We can relate the norms of the mode wavefunctions to
the Green's function by (see for example \cite{super1d})
\be
\sum_{\chi=+, -}{\psi_k^\chi}(x)^2 =2k \Im G(x, x, k) \,,
\ee
where $G (x, x', k)$ is the Green's function, which satisfies
\be
-G''(x,x',k) +V(x)G(x,x',k) - k^2G(x,x',k) =\delta (x-x')
\ee
and has only outgoing waves ($\sim e^{ik|x|}$) at infinity.

The Green's function has the symmetry property $G (x, x ', k) = G (x,
x ', -k^\ast)^\ast$, so for real $k$ we can write
\be
\sum_{\chi= +, -}\psi_k^\chi(x)^2 ={k\over i}G(x, x, k)-{k\over i}
G (x, x, -k) \,.
\ee
Thus we can compute the energy by extending the range of integration
to $-\infty$, and using $G^{(0)}(x,x,k)=i/(2k)$,
\be
\langle{\cal H}\rangle_{\text{ren}} =
-\frac{\Gamma\left(-\frac{n+1}{2}\right)}{(4\pi)^{\frac{n+3}{2}}}
\int_{-\infty}^\infty
\omega^{n+1} \left[\frac{2k}{i}G (x, x, k) - 1-{V(x)\over 2k^2}
+{n+1\over 2}\frac{k}{i\omega^2}{d^2\over dx^2}G(x, x, k)\right] \, dk \,,
\label{eqn:contourextend}
\ee
where, since we are taking the massless limit, $\omega=\sqrt{k^2}$.

Next we would like to convert this expression into a contour integral
by closing the contour at infinity in the upper half plane.  The
contour at infinity does not contribute, because for large, positive
$\Im k$,
\be
{2k\over i}G(x, x, k) \to  1+{V(x)\over 2k^2} +O (k^{-4}) \,.
\ee
Singularities in the Green's function in the upper half plane
correspond to normalizable eigenfunctions of the Hamiltonian, which
represent bound states.  Since the Hamiltonian is Hermitian, the bound
states must have real energies, so the singularities must lie on the
imaginary axis and have $\Im k < \mu$ where $\mu$ is the mass.  In this
example, we have a repulsive potential and a massless particle, each of
which is sufficient to ensure that there are no bound states at all.  Thus
the Green's function has no singularities for $\Im k\ge0$, and the only
contribution to the integral comes from the branch cut along the positive
imaginary axis coming from $\omega^{n+1}$.  Integrating around the branch
cut and using Eqs.\ (\ref{id1}) and (\ref{id2}) below gives
\be
\langle{\cal H}\rangle_{\text{ren}} =
-\frac{1}{2(4\pi)^{\frac{n+1}{2}}\Gamma\left(\frac{n+3}{2}\right)}
\int_0^\infty \kappa^{n+1} \left[2\kappa G (x, x, k) - 1 + {V(x)\over
2\kappa^2} +{n+1\over 2\kappa}{d^2\over dx^2}G(x, x, k)\right] \, d\kappa 
\ee
and then setting $n=1$ gives
\be\label{eqn:m=n=1special}
\langle{\cal H}\rangle_{\text{ren}} = -\frac{1}{8\pi}\int_0^\infty
d\kappa \,\left[ 2\kappa^3  G(x, x, i\kappa) - \kappa^2
+{V(x)\over 2}-\kappa{d^2\over dx^2}G(x, x, i\kappa)\right] \,.
\label{eqn:greenn=1}
\ee
Once one has computed the Green's function, this integral is
straightforward, though it may be necessary to resort to numerical
techniques.  We show the calculation for some example potentials in
Sec.\ \ref{sec:examples1} below.

\section{Calculational Method}

\label{sec:method}

\subsection{Model}

We will now consider the more general case of a real scalar field of
mass $\mu$ in the background of a potential that is
spherically symmetric in $m$ nontrivial coordinates, which we label by
$x$, and independent of the remaining $n$ trivial coordinates, which
we label by $y$.  The energy density is
\be
{\cal H} = \frac{1}{2} \left(\dot \phi^2 + (\nabla\phi)^2 + V(r)
\phi^2 + \mu^2 \phi^2 \right)
=  \frac{1}{2} \left(\dot \phi^2 + \frac{1}{2}\nabla^2(\phi^2)
- \phi \nabla^2 \phi + V(r)\phi^2 + \mu^2 \phi^2 \right) \,,
\ee
where $r=|x|$.  Decomposing the quantum field $\phi$ in terms of modes gives
\bea
\phi(r,\Omega,t) &=& \sum_{\ell,\ell_z}
\sqrt{\frac{2 \pi^{\frac{m}{2}}}{\Gamma\left(\frac{m}{2}\right)}}
\int\frac{d^n p}{(2\pi)^{n/2}} \frac{1}{\sqrt{2}} \cr
&& \times \left( \sum_j \frac{1}{\sqrt{\omega_j^\ell}}
\left( \psi^\ell_j(r)^\ast Y^m_{\ell\ell_z}(\Omega)^\ast e^{-ipy}
e^{i\omega_j^\ell(p) t} a_{j,p}^{\ell\ell_z}{}^\dagger  + 
\psi^\ell_j(r) Y^m_{\ell\ell_z}(\Omega) e^{ipy} 
e^{-i\omega_j^\ell(p) t} a_{j,p}^{\ell\ell_z} \right)
\right. \cr  &&+ \left.
\int_0^\infty \frac{dk}{\sqrt{\pi \omega(p)}} \left(
\psi^\ell_k(r)^\ast Y^m_{\ell\ell_z}(\Omega)^\ast e^{-ipy} 
e^{i\omega(p) t} a_{k,p}^{\ell\ell_z}{}^\dagger +
\psi^\ell_k(r) Y^m_{\ell\ell_z}(\Omega) e^{ipy}
 e^{-i\omega(p) t} a_{k,p}^{\ell\ell_z}\right) \right) \,,
\eea
where $\omega(p) = \sqrt{k^2+p^2+\mu^2}$, the bound state energies are
$\omega_j^\ell(p) = \sqrt{p^2 + \mu^2 - \kappa_j^\ell{}^2}$, and the sum
over $\ell$ gives the partial wave expansion in the $m$ nontrivial
dimensions.

The degeneracy factor $D^m_\ell$ in each partial wave is given by the
dimension of the space of symmetric tensors with $\ell$ indices, each
running from $1$ to $m$, with all traces (contractions) removed \cite{LL}. 
By the symmetry of the indices, this dimension is given by the number of
ways to make $\ell$ indices out of 0 or more 1's, 0 or more 2's, and so on,
which is the number of distinct ways to place $m-1$ dividers into
$m+\ell-1$ slots.  Removing all the traces requires subtracting the same
quantity with $\ell$ replaced by $\ell-2$.  We thus obtain
\be
D^m_\ell
= \frac{(m+\ell-1)!}{\ell!(m-1)!} - \frac{(m+\ell-3)!}{(\ell-2)!(m-1)!}
= \frac{\Gamma(m+\ell-2)}{\Gamma(m-1)\Gamma(\ell+1)}(m+2\ell-2) \,.
\ee
The wavefunctions $\psi^\ell_k(r)$ are the eigenstates of the
time-independent radial Schr\"odinger equation
\be
\left(-\frac{d^2}{dr^2} - \frac{m-1}{r}\frac{d}{dr} +
\frac{\ell(\ell+m-2)}{r^{2}} + V(r)\right) \psi^\ell_k(r) 
= k^2 \psi^\ell_k(r) \,,
\ee
which in general comprise both bound and scattering states.
The wavefunctions and creation and annihilation operators are
normalized as follows.  For the spherical harmonics,
\be
\int Y^m_{\ell\ell_z}(\Omega)^\ast Y^m_{\ell'\ell'_z}(\Omega) \, d\Omega =
\delta_{\ell \ell'} \delta_{\ell_z \ell'_z} \,,
\ee
for continuum states,
\bea
\frac{2 \pi^{\frac{m}{2}}}{\Gamma\left(\frac{m}{2}\right)}
\int_0^\infty r^{m-1} \psi^\ell_k(r)^\ast \psi^\ell_{k'}(r) \, dr &=&
\pi \delta(k-k') \cr
[a_{k,p}^{\ell\ell_z}{}^\dagger, a_{k',p'}^{\ell'\ell_z'}{}^\dagger] =
[a_{k,p}^{\ell\ell_z}, a_{k',p}^{\ell'\ell_z'}] = 0  &\qquad&
[a_{k,p}^{\ell\ell_z}, a_{k',p}^{\ell'\ell_z'}{}^\dagger] = \delta(k-k')
\delta(p-p') \delta_{\ell \ell'} \delta_{\ell_z \ell_z'}\,,
\eea
and for bound states
\bea
\frac{2 \pi^{\frac{m}{2}}}{\Gamma\left(\frac{m}{2}\right)}
\int_0^\infty r^{m-1} \psi^\ell_{j}(r) \psi^{\ell'}_{j'}(r) \, dr &=&
\delta_{j j'} \delta_{\ell \ell'} \cr
[a_{j,p}^{\ell\ell_z}{}^\dagger, a_{j',p'}^{\ell'\ell_z'}{}^\dagger] =
[a_{j,p}^{\ell\ell_z}, a_{j',p'}^{\ell'\ell_z'}] = 0 &\qquad&
[a_{j,p}^{\ell\ell_z}, a_{j',p'}^{\ell'\ell_z'}{}^\dagger] =
\delta_{j j'} \delta(p-p') \delta_{\ell \ell'} \delta_{\ell_z \ell_z'} \,.
\eea
Using these expressions, we obtain the vacuum expectation value of the
Hamiltonian,
\bea
\langle {\cal H} \rangle &=& 
\frac{1}{2}\sum_{\ell} D^m_\ell \int \frac{d^n p}{(2\pi)^n}  \left[
\sum_j  \omega_j^\ell(p) |\psi^\ell_j(r)|^2  +
\int_0^\infty \frac{dk}{\pi} \omega(p) |\psi^\ell_k(r)|^2
\right. \cr &&+ \left.
\frac{1}{4} D^2_r \left(
\sum_j  \frac{1}{\omega_j^\ell(p)} |\psi^\ell_j(r)|^2  +
\int_0^\infty \frac{dk}{\pi} \frac{1}{\omega(p)}
|\psi^\ell_k(r)|^2 \right) \right] \,,
\eea
where $D^2_r = \frac{d^2}{dr^2} + \frac{m-1}{r} \frac{d}{dr}$ is the radial
Laplacian.

\subsection{Renormalization with one subtraction}

\label{sec:onedirectsub}

For positive integer $m$ and $n$, this quantity diverges, as we expect
since we have not yet included the contribution of the counterterms.
Therefore we will calculate the result using analytic continuation in
$m$ and $n$ from values where it is convergent.  After introducing
counterterms also depending on $m$ and $n$, we will then let the
dimensions go to their physical values while holding the
renormalization conditions fixed.

The first counterterm we will introduce renormalizes the cosmological
constant.  It is simply an overall constant in the Hamiltonian, and is
fixed by the renormalization condition that the energy density of the
trivial background $V(r)=0$ is zero.  The free wavefunctions are given by
\be
\psi_k^\ell{}^{(0)}(r) = \sqrt{\pi k
\frac{\Gamma\left(\frac{m}{2}\right)}{2 \pi^{\frac{m}{2}}}}
\frac{1} {r^{\frac{m}{2}-1}} J_{\frac{m}{2}+\ell-1}(kr) \,,
\ee
which, by the Bessel function identity
\be
\sum_{\ell=0}^{\infty}
\frac{(2q+2\ell)\Gamma(2q+\ell)}{\Gamma(\ell+1)}J_{q+\ell}(z)^2 =
\frac{\Gamma(2q+1)}{\Gamma(q+1)^2}
\left(\frac{z}{2}\right)^{2q} \,,
\label{BesselId}
\ee
 satisfy the completeness relation
\be
\sum_\ell D^m_\ell |\psi_k^\ell{}^{(0)}(r)|^2 = 
\frac{1}{(4 \pi)^{\frac{m}{2}-1}}
\frac{k^{m-1}}{2\Gamma\left(\frac{m}{2}\right)} \,,
\label{eqn:complete}
\ee
independent of $r$.  Subtracting the energy in the trivial background
we have
\bea
\langle {\cal H} \rangle - \langle {\cal H} \rangle_0 &=& 
\frac{1}{2} \sum_{\ell} D^m_\ell \int\frac{d^n p}{(2\pi)^n}  \left[
\sum_j  \omega_j^\ell(p) |\psi^\ell_j(r)|^2 +
\int_0^\infty \frac{dk}{\pi} \omega(p)
\left( |\psi^\ell_k(r)|^2 - |\psi^\ell_k{}^{(0)}(r)|^2\right)
\right. \cr &&+ \left.
\frac{1}{4} D^2_r \left(
\sum_j  \frac{1}{\omega_j^\ell(p)} |\psi^\ell_j(r)|^2  +
\int_0^\infty \frac{dk}{\pi} \frac{1}{\omega(p)}
|\psi^\ell_k(r)|^2 \right) \right] \, .
\label{eqn:nolevsub}
\eea

By completeness, in each partial wave we have
\be
\sum_j |\psi^\ell_j(r)|^2 +
\int_0^\infty \frac{dk}{\pi} \left(|\psi^\ell_k(r)|^2
- |\psi^\ell_k{}^{(0)}(r)|^2 \right) = 0 \,,
\ee
so we can implement a local version of the Levinson's theorem subtraction used
in \cite{domainwall}, giving
\bea
\langle {\cal H} \rangle - \langle {\cal H} \rangle_0 &=& 
\frac{1}{2} \sum_{\ell} D^m_\ell \int\frac{d^n p}{(2\pi)^n}  \left[
\sum_j  (\omega_j^\ell(p) - \sqrt{p^2 + \mu^2}) |\psi^\ell_j(r)|^2
\right. \cr &&+ \left.
\int_0^\infty \frac{dk}{\pi} (\omega(p) - \sqrt{p^2 + \mu^2})
\left( |\psi^\ell_k(r)|^2 -  |\psi^\ell_k{}^{(0)}(r)|^2 \right)
\right. \cr &&+ \left.
\frac{1}{4} D^2_r \left(
\sum_j  \frac{1}{\omega_j^\ell(p)} |\psi^\ell_j(r)|^2 +
\int_0^\infty \frac{dk}{\pi} \frac{1}{\omega(p)}
|\psi^\ell_k(r)|^2 \right) \right] \,.
\label{eqn:yeslevsub}
\eea
This subtraction is necessary to avoid the appearance of spurious infrared
singularities in calculations in one space dimension.  These singularities
also appear in dimensions less than one,  which we will need to consider as
part of the dimensional regularization process.

Next we carry out the $p$ integral, using
\be
\int\frac{d^n p}{(2\pi)^n} \left(\sqrt{p^2 + q^2}\right)^a =
\frac{2}{\Gamma\left(\frac{n}{2}\right)(4\pi)^{n/2}}
\int_0^\infty p^{n-1} dp \, \left(\sqrt{p^2 + q^2}\right)^a =
\frac{\Gamma\left(-\frac{n+a}{2}\right)q^{n+a}}
{\Gamma\left(-\frac{a}{2}\right)(4\pi)^{n/2}} \,,
\ee
where we have done the integral by analytic continuation from values
of $a$ and $n$ where it converges.  We thus obtain
\bea
\langle {\cal H} \rangle - \langle {\cal H} \rangle_0 &=& 
-\frac{\Gamma\left(-\frac{n+1}{2}\right)}{2(4\pi)^{\frac{n+1}{2}}}
\sum_{\ell} D^m_\ell \left[
\sum_j ((\omega_j^\ell)^{n+1} -
\mu^{n+1}) |\psi^\ell_j(r)|^2 
\right. \cr &&+ \left.
\int_0^\infty \frac{dk}{\pi} (\omega^{n+1} - \mu^{n+1})
\left( |\psi^\ell_k(r)|^2 - |\psi^\ell_k{}^{(0)}(r)|^2 \right)
\right. \cr &&+ \left.
\frac{n+1}{4} D^2_r \left(
\sum_j (\omega_j^\ell)^{n-1} |\psi^\ell_j(r)|^2 +
\int_0^\infty \frac{dk}{\pi} \omega^{n-1}|\psi^\ell_k(r)|^2
\right) \right] \,,
\eea
where $\omega = \sqrt{k^2 + \mu^2}$ for the scattering states and
$\omega_j^\ell = \sqrt{\mu^2 - \kappa_j^\ell{}^2}$ for the bound states
with $k=i\kappa_j^\ell$.

Next, we must include the contribution of the counterterm
proportional to $V(r)$, which is introduced to cancel the tadpole
graph.  In dimensional regularization, the contribution to the
Hamiltonian from this counterterm is
\be
{\cal H}_1 =
\frac{\Gamma\left(\frac{1 - n - m}{2}\right)}{2(4\pi)^{\frac{m+n+1}{2}}}
 \mu^{m+n-1} V(r) \,,
\ee
so by using
\be
\int_0^\infty \frac{dk}{\pi} (\omega^{n+1}-\mu^{n+1}) k^{m-3} =
\mu^{m+n-1}\frac{\Gamma\left(\frac{m-2}{2}\right)
\Gamma\left(\frac{1-m-n}{2}\right)}
{2\pi \Gamma\left(-\frac{n+1}{2}\right)}
\ee
and Eq.\ (\ref{eqn:complete}) we have
\be
{\cal H}_1 = -\frac{\Gamma\left(-\frac{n+1}{2}\right)}{2(4\pi)^{\frac{n+1}{2}}}
\sum_{\ell} D^m_\ell\int_0^\infty \frac{dk}{\pi}
(\omega^{n+1}- \mu^{n+1})(2-m)\frac{V(r)}{2k^2} |\psi^\ell_k{}^{(0)}(r)|^2 
\ee
so that
\bea
\langle{\cal H}\rangle_{\text{ren}} &\equiv&
\langle {\cal H} \rangle - \langle {\cal H} \rangle_0 
- \langle {\cal H}_1 \rangle \cr
&=& -\frac{\Gamma\left(-\frac{n+1}{2}\right)}{2(4\pi)^{\frac{n+1}{2}}}
\sum_{\ell} D^m_\ell \left[
\sum_j ((\omega_j^\ell)^{n+1} - \mu^{n+1}) |\psi^\ell_j(r)|^2
\right. \cr &&+ \left.
\int_0^\infty \frac{dk}{\pi} (\omega^{n+1} - \mu^{n+1})
\left( |\psi^\ell_k(r)|^2  -
|\psi^\ell_k{}^{(0)}(r)|^2
\left(1 + (2-m)\frac{V(r)}{2k^2}\right) \right)
\right. \cr &&+ \left.
\frac{n+1}{4} D^2_r \left(
\sum_j (\omega_j^\ell)^{n-1}|\psi^\ell_j(r)|^2 +
\int_0^\infty \frac{dk}{\pi} \omega^{n-1}
|\psi^\ell_k(r)|^2 \right) \right]
\label{eqn:e1sub} \,.
\eea
We will use
\be
|\psi^\ell_k(r)|^2 = 2 k \Im G_\ell(r,r,k) \,,
\label{eqn:normtogreen}
\ee
where the Green's function is defined by
\be
-D_r^2 G_\ell(r,r',k) + \left(V(r) + \frac{\ell(\ell+m-2)}{r^2} - k^2\right)
G_\ell(r,r',k) = \delta^{(m)}(r-r')
\ee with the boundary conditions that it is regular at the origin and and
has only outgoing waves ($\sim e^{ikr}$) at infinity.  Using
$G(r,r,-k)=G(r,r,k)^\ast$, we can rewrite Eq.\ (\ref{eqn:e1sub}) as
\bea
\langle{\cal H}\rangle_{\text{ren}} &\equiv&
\langle {\cal H} \rangle - \langle {\cal H} \rangle_0 
- \langle {\cal H}_1 \rangle \cr
&=& -\frac{\Gamma\left(-\frac{n+1}{2}\right)}{2(4\pi)^{\frac{n+1}{2}}}
\sum_{\ell} D^m_\ell \left[
\sum_j ((\omega_j^\ell)^{n+1} - \mu^{n+1}) |\psi^\ell_j(r)|^2
\right. \cr &&+ \left.
\int_{-\infty}^\infty \frac{dk}{\pi} (\omega^{n+1} - \mu^{n+1})
\frac{k}{i} \left(  G_\ell(r,r,k) - G_\ell^{(0)}(r,r,k)
\left(1 + (2-m)\frac{V(r)}{2k^2}\right) \right)
\right. \cr &&+ \left.
\frac{n+1}{4} D^2_r \left(
\sum_j (\omega_j^\ell)^{n-1} |\psi^\ell_j(r)|^2 + 
\int_{-\infty}^\infty \frac{dk}{\pi} \omega^{n-1}
\frac{k}{i} G_\ell(r,r,k) \right) \right]
\label{eqn:e1subGreen} \,.
\eea

These subtractions are sufficient to render the theory finite for
$m+n<3$.  However, it appears that if we set $n=1$, the gamma function
will cause Eq.\ (\ref{eqn:e1subGreen}) to diverge.  In fact, as we
will see in Appendix\ \ref{sec:localsum}, it does not, because the quantity
in brackets vanishes.  But here we will keep $n$ general and instead close
the contour of integration in the upper half $k$ plane.  For sufficiently
small $n$ the contour at infinity does not contribute.  There is a pole for
each bound state at $k=iE_j$, where $E_j<\mu$ is the bound state energy,
and the contributions from these poles exactly cancel the sum over bound
states \cite{Bordag} in Eq.\ (\ref{eqn:e1subGreen}). Thus the final result
is just given by the contribution from the branch cut along the imaginary
axis from $\mu$ to $\infty$  resulting from $\omega^{n+1}$, which
contributes 
\be
\Omega^{n+1} \left(i^{n+1} - (-i)^{n+1}\right) = 
2i\Omega^{n+1} \sin \frac{(n+1)\pi}{2} \,,
\label{id1}
\ee
where $\Omega = \sqrt{\kappa^2 - \mu^2}$ and $k=i\kappa$.  Then using
the identity
\be
\sin \pi z  = -\frac{\pi}{\Gamma\left(z+1\right) \Gamma\left(-z\right)}
\label{id2}
\ee
we have
\bea
\langle{\cal H}\rangle_{\text{ren}} &\equiv&
\langle {\cal H} \rangle - \langle {\cal H} \rangle_0 
- \langle {\cal H}_1 \rangle \cr
&=& -\frac{1}{2(4\pi)^{\frac{n+1}{2}}\Gamma\left(\frac{n+3}{2}\right)}
\sum_{\ell} D^m_\ell \int_\mu^\infty d\kappa \, 2\kappa \Omega^{n+1} \left[  
G_\ell(r,r, i\kappa) - G_\ell^{(0)}(r,r, i\kappa)
\phantom{\frac{1}{1}}\right. \cr && \left. \times
\left(1 - (2-m)\frac{V(r)}{2\kappa^2}\right)
-\frac{n+1}{4 \Omega^{2}} D^2_r G_\ell(r,r, i\kappa) \right] \,.
\label{eqn:e1subimag}
\eea
We can now put in integer values of $m$ and $n$ without any
divergence, as long as $m+n<3$.  Eq.\ (\ref{eqn:e1subimag}) can also be
efficiently evaluated numerically \cite{MITgroup}.

\subsection{Higher subtractions}

When we have $m+n=3$ space dimensions, we will need to introduce a
second counterterm, $\frac{1}{2} c V(x)^2$.  The first subtraction
is particularly easy to define because there is a natural scheme,
specified by the complete cancellation of the tadpole graph.  Higher
subtractions require a definition in terms of a renormalization scale,
which can be chosen arbitrarily.  In choosing this scale, we must be
able to relate it to physical inputs, such as masses and coupling
constants, in order to define a predictive theory.

To define the counterterm precisely, we consider the two-point
function $\Pi(p^2)$ in dimensional regularization.  It diverges as we
approach the physical dimension.  The divergence is canceled by the
contribution of the counterterm to the two-point function, which is
just $c$.  We define the renormalization scale $M$ by taking
$c=-\Pi(M^2)$.  With this definition, we have
\bea
c &=& \frac{i}{2} \int_0^1 d\lambda 
\int \frac{dE}{2\pi} \frac{d^{n+m} q}{(2\pi)^{n+m}}
\frac{1}{(E^2 - q^2 - \mu^2 + M^2 \lambda (1-\lambda) + i\epsilon)^2} \cr
&=& \frac{1}{2(4\pi)^{\frac{n+m}{2}} \Gamma\left(\frac{n+m}{2}\right)}
\int_0^\infty \frac{q^{n+m-1}}{\omega (4\omega^2-M^2)} \, dq \,,
\label{eqn:cterm}
\eea
where $q$ is the total momentum and we have integrated over the
Feynman parameter $\lambda$ and the loop energy $E$.  Typically we will
choose $M^2 = \mu^2$, except in the case of massless theories, where to
avoid infrared singularities we will choose a spacelike renormalization
point $M^2 < 0$.

This regulated expression is defined precisely as an analytic function of
the dimension.  Our goal is now to rewrite it in a way that allows us to
incorporate it into our expression for the energy, Eq.\
(\ref{eqn:e1subimag}), which is also given as an analytic function of the
dimension.  We express Eq.\ (\ref{eqn:complete}) in terms of Green's
functions and analytically continue to express Eq.\ (\ref{eqn:cterm}) as
\bea
c &=& \frac{1}{2(4\pi)^{\frac{n+1}{2}}\Gamma\left(\frac{n+3}{2}\right)}
\int_\mu^\infty \Omega^{n+1} f(\kappa,M) 
\frac{1}{(4 \pi)^{\frac{m}{2}-1}}
\frac{\kappa^{m-1}}{2\Gamma\left(\frac{m}{2}\right)} \, d\kappa \cr 
&=& \frac{1}{2(4\pi)^{\frac{n+1}{2}}\Gamma\left(\frac{n+3}{2}\right)}
\sum_\ell D^m_\ell \int_\mu^\infty \Omega^{n+1} f(\kappa,M) 
2\kappa G_\ell^{(0)}(r,r, i\kappa) \, d\kappa \,,
\label{eqn:ctermrot}
\eea
where $f(\kappa,M)$ is given in terms of the hypergeometric function as
\be
f(\kappa,M) =
\frac{2 (m-4)(m-2)(2\kappa)^{2 - m}}{(4\kappa^2 - M^2)^{3-\frac{m}{2}}
\sin \left(\frac{m\pi}{2}\right)} \,\,
{}_2 F_1\left(\frac{1}{2},3 -
\frac{m}{2},\frac{3}{2},\frac{M^2}{M^2 - 4\kappa^2}\right)
\label{eqn:fkappa}
\ee
as we show in Appendix\ \ref{sec:fderivation}.

Since we will eventually take the limit where $m$ becomes an integer,
we note that
\bea
f(\kappa, M) = \frac{12 \kappa^2-M^2}{2\kappa^2(4\kappa^2 - M^2)^2}
&\qquad& \mbox{for $m=1$,} \cr
f(\kappa, M) = \frac{1}{\pi\kappa^2(4\kappa^2 - M^2)}
\left(1 + \frac{4\kappa^2 \arctan\frac{M}{\sqrt{4\kappa^2 - M^2}}}
{M\sqrt{4\kappa^2 - M^2}}\right)
&\qquad& \mbox{for $m \to 2$, and} \cr
f(\kappa, M) = \frac{1}{2\kappa^2(4\kappa^2-M^2)}
&\qquad& \mbox{for $m=3.$}
\label{eqn:fkappa1}
\eea
Eq.\ (\ref{eqn:ctermrot}) is now in a form where we can include it
under the integral sign in Eq.\ (\ref{eqn:e1subimag}) and obtain 
\bea
\langle{\cal H}\rangle_{\text{ren}} &\equiv&
\langle {\cal H} \rangle - \langle {\cal H} \rangle_0 
- \langle {\cal H}_1 \rangle - \langle {\cal H}_2 \rangle \cr
&=& -\frac{1}{2(4\pi)^{\frac{n+1}{2}}\Gamma\left(\frac{n+3}{2}\right)}
\sum_{\ell} D^m_\ell
\int_\mu^\infty d\kappa \, 2\kappa \left[ \Omega^{n+1} \left(
G_\ell(r,r, i\kappa) - G_\ell^{(0)}(r,r, i\kappa)
\phantom{\frac{1}{1}} \right. \right. \cr && \left. \left. \times
\left(1 - \frac{V(r)}{2\kappa^2} (2-m) + V(r)^2 f(\kappa,M) \right) \right)
- \frac{n+1}{4}\Omega^{n-1} D^2_r 
G_\ell(r,r, i\kappa) \right] \,.
\label{eqn:e2subimag}
\eea

Before we can take the limit where $m+n=3$, however, there is one more
potential divergence in Eq.\ (\ref{eqn:e2subimag}).  Our subtraction has
cancelled the terms of order $1$, $V(r)/\kappa^2$ and $V(r)/\kappa^4$ in
the large-$\kappa$ expansion of the norm of the wavefunctions.  But there
could also be a term of order $D_r^2 V(r)/\kappa^4$, which will generate a
divergence in this case.  In the renormalization of the composite
operator $T_{\mu\nu}$, we have a renormalization counterterm
$\frac{c'}{2} \frac{m+n-1}{4(m+n)}
(\partial_\mu \partial_\nu - g_{\mu\nu} \partial_\lambda
\partial^\lambda)\phi^2$ \cite{Collins}.  Since we are considering
just $T_{00}$ here, this counterterm becomes $\frac{c'}{2}
\frac{m+n-1}{4(m+n)} \nabla^2 \phi^2$, exactly the form needed to cancel
the remaining divergence.\footnote{If we had chosen conformal instead of
minimal coupling for the fields, which corresponds to adding the extra term
$\frac{1}{2} \frac{m+n-1}{4(m+n)} (\partial_\mu \partial_\nu - g_{\mu\nu}
\partial_\lambda \partial^\lambda)\phi^2$ in the original Lagrangian, the
divergent term would have cancelled automatically between the bulk term and
the surface term and no renormalization would be necessary.
However, conformally coupled theories have {\em
classical} violations of the energy conditions \cite{Barcelo:2000zf}.}
We fix this counterterm by subtracting the tadpole diagram with the
composite operator carrying momentum $p^2 = {M'}^2$. Aside from this
change, it is analogous to the tadpole subtraction above, with
$\Omega^2 V(r)$ replaced by $D_r^2 V(r)$.  The scale $M'$ is then
specified through the renormalization condition on the composite
operator (and would typically be chosen equal to $M$). As with $M$, a
massless theory will require spacelike ${M'}^2 < 0$, while in a
massive theory we may set $M' = \mu$.  Thus we obtain the contribution
\be
\langle {\cal H}_{2'} \rangle 
= -\frac{(m+n-1)}{4(n+1)}
\frac{1}{2(4\pi)^{\frac{n+1}{2}}\Gamma\left(\frac{n+1}{2}\right)}
\sum_{\ell} D^m_\ell
\int_\mu^\infty d\kappa \, \Omega^{n-1} 2\kappa
G_\ell^{(0)}(r,r, i\kappa)
\frac{D_r^2 V(r)}{\kappa^2-{M'}^2}(2-m) \,.
\ee
This term is a total derivative, so it does not contribute to the total
energy.  We can split the contribution of this term beween the bulk and
derivative terms so that it renders them both separately finite at
integer dimensions, giving
\bea
\langle{\cal H}\rangle_{\text{ren}}
&=& -\frac{1}{2(4\pi)^{\frac{n+1}{2}}\Gamma\left(\frac{n+3}{2}\right)}
\sum_{\ell} D^m_\ell
\int_\mu^\infty d\kappa \, 2\kappa \left[ \Omega^{n+1} 
\left(G_\ell(r,r, i\kappa) -  G_\ell^{(0)}(r,r, i\kappa)
\phantom{\frac{1}{1}} \right. \right. \cr && \left. \left. \times
\left(1 - \frac{V(r)}{2\kappa^2} (2-m) +  V(r)^2 f(\kappa,M)
- \frac{D_r^2 V(r)}{8(\kappa^2-{M'}^2) \Omega^2}(2-m)^2 
\right) \right)
\right. \cr &&- \left.
\frac{n+1}{4}\Omega^{n-1}
\left( D^2_r G_\ell(r,r, i\kappa) + \frac{ D^2_r V(r)}{2(\kappa^2-{M'}^2)}
(2-m) G_\ell^{(0)}(r,r, i\kappa) \right) \right] \,.
\eea
By Eq.\ (\ref{eqn:complete}), the sum over $\ell$ of the free Green's
function weighted by the degeneracy factor is independent of $r$, so we can
pull the derivative outside in the last line, giving
\bea
\langle{\cal H}\rangle_{\text{ren}}
&=& -\frac{1}{2(4\pi)^{\frac{n+1}{2}}\Gamma\left(\frac{n+3}{2}\right)}
\sum_{\ell} D^m_\ell
\int_\mu^\infty d\kappa \, 2\kappa \left[ \Omega^{n+1} 
\left(G_\ell(r,r, i\kappa) -  G_\ell^{(0)}(r,r, i\kappa)
\phantom{\frac{1}{1}}
\right. \right. \cr && \left. \left. \times
\left(1 - \frac{V(r)}{2\kappa^2} (2-m) +  V(r)^2 f(\kappa,M)
- \frac{D_r^2 V(r)}{8(\kappa^2-{M'}^2) \Omega^2}(2-m)^2
\right) \right)
\right. \cr &&- \left.
\frac{n+1}{4}\Omega^{n-1} D^2_r
\left(G_\ell(r,r, i\kappa) + \frac{V(r)}{2(\kappa^2-{M'}^2)} (2-m)
G_\ell^{(0)}(r,r, i\kappa) \right) \right] \,.
\eea

In this form, we see explicitly that the subtraction has cancelled the
leading terms for large $\kappa$:  In the surface term, inside the
derivative we have implemented the same subtraction of the leading behavior
of the Green's function as we found for the tadpole graph in the bulk
term; in the bulk term, we have subtracted the leading term proportional
to $D_r^2 V(r)$ (which is derived explicitly for $m=1$ in Eq.\
(\ref{eqn:asymp1d})).  For the purposes of calculation, however, it is
often easier to work with the combined expression
\bea
\langle{\cal H}\rangle_{\text{ren}}
&=& -\frac{1}{(4\pi)^{\frac{n+1}{2}}\Gamma\left(\frac{n+3}{2}\right)}
\sum_{\ell} D^m_\ell
\int_\mu^\infty \kappa \Omega^{n+1} \left[ 
\left(1 - \frac{n+1}{4\Omega^2} D^2_r\right)
G_\ell(r,r, i\kappa) -  G_\ell^{(0)}(r,r, i\kappa)
\phantom{\frac{1}{1}} \right. \cr && \left. \times
\left(1 - \frac{(2-m)V(r)}{2\kappa^2} +  V(r)^2 f(\kappa,M)
+ \frac{(2-m) D_r^2 V(r)}{8(\kappa^2-{M'}^2) \Omega^2}(m+n-1)
\right)\right] \, d\kappa \,.
\label{e2subimag2}
\eea

\section{Examples with one relevant dimension and one irrelevant dimension}

\label{sec:examples1}

\subsection{The general case}

To illustrate our method, we would like to carry out some
sample calculations in the case of $m=1$ and $n=1$.  Since we are in
$2+1$ dimensions, we only need one subtraction, proportional to $V(x)$.
For $m=1$, the sum over partial waves reduces to a sum over the symmetric
and antisymmetric channels.  The free wavefunctions thus become
\be
\psi^+_k {}^{(0)}(x) = \cos kx \qquad \psi^-_k {}^{(0)}(x) = \sin kx \,,
\ee
so that
\be
|\psi^+_k{}^{(0)}(x)|^2 + |\psi^-_k{}^{(0)}(x)|^2 = 1\,.
\ee
We can sum over the two modes to get the overall Green's function,
\be
G(x,x,k)=G^+(x,x,k)+G^-(x,x,k) \,,
\ee
with
\be
G^{(0)}(x,x,k)=\frac{i}{2k}\,.
\ee
Then from Eq.\ (\ref{eqn:e1subimag}) we have
\be
\label{eqn:m=1}
\langle{\cal H}\rangle_{\text{ren}}
= -\frac{1}{2(4\pi)^{\frac{n+1}{2}}\Gamma\left(\frac{n+3}{2}\right)}
\int_\mu^\infty d\kappa \, \Omega^{n+1} \left[
2 \kappa G(r,r, i\kappa) - 1+ \frac{V(r)}{2\kappa^2}
- \frac{n+1}{2\Omega^2} \kappa D^2_r G(r,r, i\kappa) \right]
\ee
and for $n=1$ we have
\be
\langle{\cal H}\rangle_{\text{ren}}
= -\frac{1}{8\pi}
\int_\mu^\infty d\kappa \, \Omega^2 \left[2\kappa G(r,r, i\kappa) - 1
+ \frac{V(r)}{2\kappa^2} - \frac{\kappa}{\Omega^2} \frac{d^2}{dx^2}
G(r,r, i\kappa) \right] \,,
\ee
which reduces to Eq.\ (\ref{eqn:m=n=1special}) when $\mu=0$.  For
simplicity, we will consider the massless case for the remainder of
this section.

\subsection{Outside a potential with compact support}
We next consider a potential that vanishes for all $|x|> a$ and
calculate the energy density in this region.  In this
case, the only counterterm is the vacuum energy, and we get
\be
\langle{\cal H}\rangle_{\text{ren}} =
-{1\over 8\pi}\int_0^\infty
d\kappa\,\left[\kappa^2\left(2\kappa G (x, x,\kappa) -1\right) 
-\kappa {d^2\over dx^2} G (x, x, \kappa)\right] \,.
\ee
The Green's function for $x, x' > a$ is
\be
G(x, x', k) ={i\over 2k}\left(e^{-ikx_<}+r(k)e^{ikx_<}\right)e^{ikx_>} \,,
\ee
where $r(k)$ is the reflection amplitude.  Thus
\be\label{eqn:Goutside}
2\kappa G (x, x, i\kappa) = 1+r(i\kappa) e^{-2\kappa x}
\ee
and
\be\label{eqn:outside}
\langle{\cal H}\rangle_{\text{ren}} =
{1\over 8\pi}\int_0^\infty d\kappa\,\kappa^2 r(i\kappa) e^{-2\kappa x}\, .
\ee
In the large $x$ limit, only small $\kappa$ contribute in the
integral.  As a result, the integral depends only on $r(0) = -1$
(at $k=0$ we always have perfect reflection\footnote{The only
exceptions to this rule are potentials with a bound state precisely at
threshold \cite{Barton,super1d}, which include reflectionless
potentials.}), so we can approximate
\be\label{eqn:outside-limit}
\langle{\cal H}\rangle_{\text{ren}} \approx
-{1\over 8\pi}\int_0^\infty d\kappa\,\kappa^2 e^{-2\kappa x}
=-{1\over 32\pi x^3}\,.
\ee

\subsection{Square barrier}
Next we consider a square barrier with $V = V_0$ for $|x|<a$ and $V=0$
otherwise.  In this case, we can compute the normal mode wave
functions in closed form, but must do a numerical integration at the
end.  Outside the barrier, the energy is given by 
Eq.\ (\ref{eqn:outside}) with
\be \label{eqn:r-square}
r = -{V_0e^{2\kappa a}\tanh2\kappa ' a\over
2\kappa\kappa'+ (\kappa^2+\kappa '^2)\tanh2\kappa ' a}
\ee
and $\kappa '^2 =\kappa^2+V_0$.

Thus, outside the barrier, the energy is
\bea
\langle{\cal H}\rangle_{\text{ren}}& = &
-{V_0\over 8\pi}\int_0^\infty
d\kappa \, {\kappa^2 e^{-2\kappa (x- a)}\tanh2\kappa ' a\over
2\kappa\kappa '+ (\kappa^2+\kappa '^2)\tanh2\kappa ' a}\cr
& = &-{V_0 \over 8\pi a}\int_0^\infty
dq \, {q^2 e^{-2q (y- 1)}\tanh2q'\over
2qq '+ (q^2+q '^2)\tanh2q'} \,,
\eea
where we have defined the dimensionless quantites $y = x/a$, $q
=\kappa a$, and $q'=\sqrt{q^2+v} =\kappa' a$ where $v = V_0 a^2$.  Note
that the integrand cannot be less than 0, so the energy outside the barrier
is always negative.

Far from the potential, specifically where $y-1\gg1/\sqrt{v}$ and
$y-1\gg1/v$, the contribution comes primarily from $q\ll\sqrt{v}$, and
thus $q '\approx\sqrt{v}$.  The integral is then
\be
\int_0^\infty{dq\, q^2e^{-2q (y-1)}\over v} ={1\over 4 (y-1)^3 v}
\ee
and
\be\label{eqn:outside-square-limit}
\langle{\cal H}\rangle_{\text{ren}} \approx-{1\over 32\pi (x-a)^3}
\ee
in agreement with Eq.\ (\ref{eqn:outside-limit}).

Close to the potential, specifically when $y-1\ll1/\sqrt{v}$ and
$y-1\ll1$, the contribution comes mostly from $q\gg\sqrt{v}$ and
$q\gg1$.  Thus $q'\approx q$ and $\tanh 2q '\approx 1$.  The integral
becomes
\be
{1\over 8(y-1)}
\ee
and
\be
\langle{\cal H}\rangle_{\text{ren}} \approx-{V_0\over 64\pi(x- a)} \,.
\ee

Inside the barrier, we have
\be
G (x, x', k) ={i\over k '} {(k '\cos k ' (x_<+a) - ik\sin k ' (x_<+a))
(k '\cos k ' (x_>-a) + ik\sin k ' (x_>-a))\over
2kk '\cos 2k ' a - i (k^2+k '^2)\sin 2k ' a} \,,
\ee
where $k' = \sqrt{k^2 - V_0}$, so we can write
\be
G (x, x, i\kappa) =
{1\over 2\kappa '}{(\kappa^2+\kappa '^2)\cosh 2\kappa ' a
+2\kappa\kappa '\sinh 2\kappa ' a+V_0\cosh 2\kappa ' x\over
2\kappa\kappa '\cosh2\kappa ' a+ (\kappa^2+\kappa '^2)\sinh2\kappa ' a} \,.
\ee
We can then split the energy into two parts,
\be
\langle{\cal H}\rangle_{\text{ren}} = E_0+E_1 (x) \,,
\ee
where $E_1$ depends on position, but $E_0$ does not.

The position-independent part is
\bea
E_0 &=&-{1\over 8\pi}\int_0^\infty d\kappa \, \bigg\{
{\kappa^3\over\kappa '}{(\kappa^2+\kappa '^2)\cosh 2\kappa ' a
+2\kappa\kappa '\sinh 2\kappa ' a \over
2\kappa\kappa '\cosh2\kappa ' a+ (\kappa^2+\kappa '^2)\sinh2\kappa '
a}- \kappa^2+{V_0\over2}\bigg\}\cr
&=& -{V_0^2\over 8\pi}\int_0^\infty d\kappa \,
\frac{1}{2 \kappa'} \frac{2\kappa + \kappa' \tanh 2\kappa' a}
{2 \kappa \kappa' + (\kappa^2 + {\kappa'}^2)\tanh 2\kappa' a} \cr
&=& -{V_0^2 a\over 8\pi}\int_0^\infty dq \,
\frac{1}{2 q'} \frac{2q + q' \tanh 2 q'} {2 q q' + (q^2 + {q'}^2)\tanh 2q'}
\eea
and is always negative.  In the limit where $v\gg1$, we can
approximate $\tanh 2q'\approx1$ to get
\be
E_0 = -{V_0^{3/2}\over 12\pi} \,.
\ee

The position-dependent part is
\bea\label{eqn:E1q}
E_1 (x) &=&{1\over 8\pi}\int_0^\infty d\kappa \,
{\kappa\over\kappa'}\,{(2\kappa'^2-\kappa^2)V_0\cosh 2\kappa' x
\over
2\kappa\kappa' \cosh 2\kappa' a+ (\kappa^2+\kappa'^2)\sinh2\kappa 'a} \cr
& = &{V_0 \over 8\pi a}\int_0^\infty {q\,dq\over q '}
{(2q'^2-q^2)\cosh 2q 'y \over 2qq '\cosh2q'+ (q^2+q '^2)\sinh2q'}
\eea
and is always positive.

Note that the the dominant term in the integrand in Eq.\
(\ref{eqn:E1q}) is suppressed by $e^{-2q ' (1-y)} < e^{-2\sqrt{v}
(1-y)}$.  Thus, far from the edge of the potential, where
$1-y \gg 1/\sqrt{v}$, $E_1$ is negligible.

Close to the edge of the potential, with
$1-y\ll1/\sqrt{v}$ and $1-y\ll1$, the integral is $1/(8(1-y))$ and
\be
E_1 (x) \approx-{V_0\over 64\pi(a-x)}
\ee
which cancels, in a principal value sense, the divergence outside the
barrier.

\begin{figure}
\begin{center}
\leavevmode\epsfbox{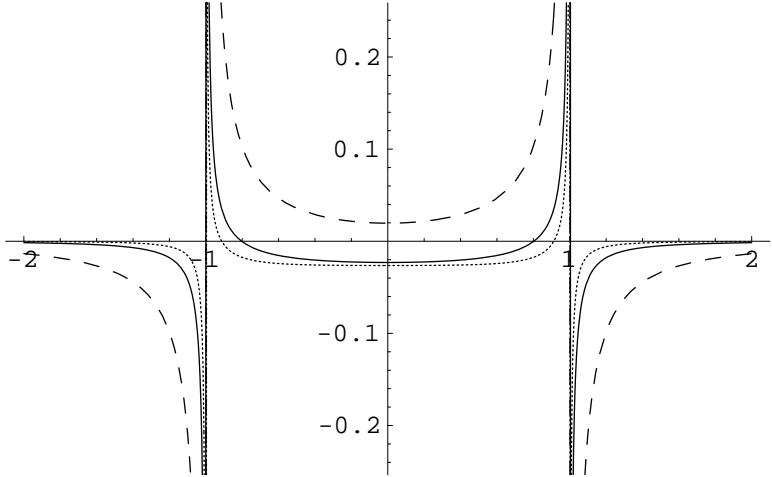}
\end{center}
\caption{Energy density in units of $V_0^{3/2}$ for the square barrier
of width 1 and heights 0.05 (dashed), 1 (solid), and 5 (dotted).  As
$V_0$ increases, the positive energy becomes concentrated more and
more near the edge of the barrier.  In units of $V_0^{3/2}$, the
outside energy decreases with $V_0$, but in absolute terms it
approaches a fixed limit given by  Eq.\ (\ref{eqn:outside-square-limit}).}
\label{fig:square}
\end{figure}

The sign of the energy density at the center of the barrier depends on the
competition between the position-dependent and position-independent
parts.  For large $v$, the position-dependent part is suppressed in
the center, and the energy density is negative.  For small $v$, it is
positive.  The total energy density is shown for several values of $v$
in Fig.\ \ref{fig:square}.

It has long been known that the energy density near a perfectly
reflecting boundary is zero if one uses the ``conformally coupled''
stress-energy tensor, but diverges if one uses the minimally coupled
one, as we have done above.  Kennedy, Critchley, and Dowker \cite{Kennedy:ar}
argue that since the total energy is the same in the two cases, there
must be a surface energy associated with the perfect conductor in the
minimal case.  Ford and Svaiter \cite{Ford:1998he} found that the
surface energy could be seen by allowing the boundary to fluctuate.

Here, we can see the situation by approximating a perfect conductor by
a square barrier with $a$ fixed and $V_0\to\infty$.  We can produce
the conformal Hamiltonian density by including half the value of the
total derivative term,
\be
{\cal H}_{\text{conformal}}
= \frac{1}{2} \dot \phi^2 + \frac{1}{8}\nabla^2(\phi^2)
- \frac{1}{2} \phi \nabla^2 \phi + V(r)\phi^2 \,.
\ee
This choice gives zero energy outside the barrier and removes the
divergence of the energy density everywhere inside.  With the
minimally coupled Hamiltonian, the energy outside goes to Eq.\
(\ref{eqn:outside-square-limit}) as $V_0$ becomes large, while the
positive energy inside clusters ever closer to the boundary, as shown
in Fig.\ \ref{fig:square}.  Since the change to the total derivative
term does not affect the total energy, we can see that the ``surface
energy'' located just inside the boundary cancels the divergent
negative energy outside.

\section{Examples with one relevant dimension and two irrelevant dimensions}

\subsection{The general case}

To carry out calculations in $3+1$ dimensions, we now need subtractions
proportional to $V(x)$, $V(x)^2$, and $V''(x)$.  We will use the
renormalization scheme defined in Section\ \ref{sec:method}.
For $m=1$, using Eq.~(\ref{eqn:fkappa1}) and evaluating Eq.\
(\ref{e2subimag2}) with $m=1$ and $n=2$ gives
\bea
\langle{\cal H}\rangle_{\text{ren}}
&=& -\frac{1}{12\pi^2}
\int_\mu^\infty d\kappa \, \Omega^3 \left[  2\kappa G(x,x, i\kappa)
- \frac{3\kappa}{2\Omega^2} \frac{d^2}{dx^2} G(x,x, i\kappa) 
- 1 + \frac{V(x)}{2\kappa^2}
\right. \cr &&- \left.
\frac{V(x)^2(12 \kappa^2 - M^2)}{2\kappa^2(4\kappa^2 - M^2)^2}
- \frac{V''(x)}{4(\kappa^2-{M'}^2) \Omega^2} \right]\,,
\label{eqn:m=1n=2}
\eea
where the Green's function has again been summed over the symmetric
and antisymmetric channels.  Again, we will
restrict our attention to massless fields for simplicity.
 
\subsection{Outside a potential with compact support}

The wave functions and Green's functions are just as in
Sec.\ \ref{sec:examples1}.  Again, since the potential vanishes, the
only counterterm is the vacuum energy.  Thus Eq.\ (\ref{eqn:m=1n=2})
reduces to
\be
\langle{\cal H}\rangle_{\text{ren}} =
-{1\over 12\pi^2}\int_0^\infty
d\kappa\,\left[\kappa^3\left(2\kappa G (x, x,\kappa) -1\right) 
-\frac{3}{2}\kappa^2 {d^2\over dx^2} G (x, x, \kappa)\right]
\ee
outside the potential, and so from Eq.\ (\ref{eqn:Goutside}),
\be
\langle{\cal H}\rangle_{\text{ren}}  =
\frac{1}{6\pi^2}
\int_0^\infty d\kappa \, \kappa^3r (i\kappa) e^{-2\kappa x} \,.
\ee
In the large $x$ limit, we can again take $r(i\kappa) \approx r(0) =
-1$, to get
\be\label{eqn:outside-limit3}
\langle{\cal H}\rangle_{\text{ren}}  =
-\frac{1}{6\pi^2}
\int_0^\infty d\kappa \, \kappa^3 e^{-2\kappa x}=-\frac{1}{16\pi^2x^4} \,,
\ee
a well-known result.

\subsection{Square barrier}

Outside a square barrier with width $a$ and height $V_0$, the
reflection coefficient is given by Eq.\ (\ref{eqn:r-square}), and the
energy is
\bea
\langle{\cal H}\rangle_{\text{ren}}& = &
-{V_0\over 6\pi^2}\int_0^\infty
d\kappa \, {\kappa^3 e^{-2\kappa (x- a)}\tanh2\kappa ' a\over
2\kappa\kappa '+ (\kappa^2+\kappa '^2)\tanh2\kappa '
a}\cr
& = &-{V_0 \over 6\pi^2 a^2}\int_0^\infty
dq \, {q^3 e^{-2q (y- 1)}\tanh2q'\over
2qq '+ (q^2+q '^2)\tanh2q'} \,.
\eea

Far from the potential, we approximate $q '\approx\sqrt{v}\gg q$.  The
integral is
\be
\int_0^\infty{dq\, q^3e^{-2q (y-1)}\over v} ={3\over 8 (y-1)^4 v}
\ee
and
\be\label{eqn:outside-square-limit3}
\langle{\cal H}\rangle_{\text{ren}} \approx-{1\over 16\pi^2 (x-a)^4} \,,
\ee
in agreement with Eq.\ (\ref{eqn:outside-limit3}).

Close to the potential, we approximate $q'\approx q$ and $\tanh 2q
'\approx 1$.  The integral becomes
\be
{1\over 16(y-1)^2}
\ee
and
\be
\langle{\cal H}\rangle_{\text{ren}} \approx-{V_0\over 96\pi^2(x- a)^2}\,.
\ee

Inside the potential, we need the renormalized form,
\be
\langle{\cal H}\rangle_{\text{ren}}  =
-\frac{1}{12\pi^2}
\int_0^\infty d\kappa\, \left[\kappa^3
\left(  2\kappa G - 1+\frac{V(x)}{2\kappa^2}
-\frac{(12\kappa^2-M^2)V(x)^2}{2\kappa^2 (4\kappa^2-M^2)^2} \right)
-\frac{3}{2} \kappa^2 \frac{d^2}{dx^2} G \right] \,,
\ee
where $M^2 < 0$ is the spacelike renormalization point.

For the square barrier, we get a position-independent part,
\bea
E_0 &=&-{1\over 12\pi^2}\int_0^\infty d\kappa\,\bigg\{
{\kappa^4\over\kappa '}{(\kappa^2+\kappa '^2)\cosh 2\kappa ' a
+2\kappa\kappa '\sinh 2\kappa ' a
\over
2\kappa\kappa '\cosh2\kappa ' a+ (\kappa^2+\kappa '^2)\sinh2\kappa ' a}\cr
& &\qquad\qquad\qquad- \kappa^3+{\kappa V_0\over2}
-\frac{\kappa(12\kappa^2+\tilde M^2)V_0^2}{2 (4\kappa^2+\tilde M^2)^2}
\bigg\}\cr
&=&-{V_0^2\over 12\pi^2}\int_0^\infty d\kappa\,\bigg\{
\frac{\kappa}{2 \kappa'} \frac{2\kappa + \kappa' \tanh 2\kappa' a}
{2 \kappa \kappa' + (\kappa^2 + {\kappa'}^2)\tanh 2\kappa' a}
-\frac{\kappa(12\kappa^2 + \tilde M^2)}{2 (4\kappa^2 + \tilde M^2)^2}
\bigg\}\cr
&=&-{V_0^2\over 12\pi^2}\int_0^\infty dq \,\bigg\{
\frac{q}{2 q'} \frac{2q + q' \tanh 2q'}
{2 q q' + (q^2 + {q'}^2)\tanh 2q'}
-\frac{q(12q^2 + t^2)}{2 (4q^2 + t^2)^2} \bigg\} \,,
\eea
where $\tilde M^2 = -M^2$ and $t = \tilde M a$.

We can isolate the dependence on the renormalization scale by using
\be
\int_0^\infty dq\bigg\{
\frac{3}{8q'}-\frac{q(12q^2+t^2)}{2 (4q^2+t^2)^2}
\bigg\}
 = \frac{3}{16}\ln\frac{t^2}{v} + \frac{1}{8}
\ee
to obtain
\be
E_0 = -{V_0^2\over 12\pi^2}
\left(\frac{3}{16}\ln\frac{\tilde M^2}{V_0} + \frac{1}{8}
+ \int_0^\infty dq \,\bigg\{
\frac{q}{2 q'} \frac{2q + q' \tanh 2q'}
{2 q q' + (q^2 + {q'}^2)\tanh 2q'}
- \frac{3}{8q'} \bigg\} \right) \,.
\ee
In the limit where $v\gg 1$, we can approximate $\tanh2q'\approx1$,
the integral gives $-7/32$, and we obtain
\be E_0 =
\frac{V_0^2}{64\pi^2}\left[\ln\frac{V_0}{\tilde M^2}+\frac{1}{2}\right] \,,
\ee
consistent with the result obtained from the effective potential
\cite{Coleman}.

The position-dependent part is
\bea\label{eqn:E1q3}
E_1 (x) &=&{1\over 12\pi^2}\int_0^\infty d\kappa\,
{\kappa^2\over\kappa'}\,{(3\kappa'^2-\kappa^2)V_0\cosh 2\kappa' x
\over
2\kappa\kappa' \cosh 2\kappa' a+ (\kappa^2+\kappa'^2)\sinh2\kappa 'a}
\cr
& = &{V_0\over 12\pi^2 a^2}\int_0^\infty dq{q^2\over q '}
{(3q'^2-q^2)\cosh 2q 'y
\over
2qq '\cosh2q'+ (q^2+q '^2)\sinh2q'}
\eea
and is always positive.

Far from the edge of the potential, $E_1$ is negligible.  Close to the
edge, where we can approximate $q' \approx q$ and $\sinh 2q  \approx \cosh
2q  \approx e^{2q}/2$, the integral becomes $1/(8(1-y)^2)$, and
\be
E_1 (x) \approx-{V_0\over 96\pi^2(a-x)^2} \,,
\ee
which cancels the divergence outside the
barrier.

These results do not reflect any contribution from the $V''(x)$
counterterm.  In this case it vanishes for all $|x|\neq a$, since the
potential is constant.  Furthermore, the contribution to the total
energy from this term is also zero, since it is a total derivative.
If we imagine that the square barrier represents the limit in which a smooth
potential gets steeper and steeper, we will find large equal and
opposite contributions to the energy localized in the tiny region
on both sides of the boundary.  As long as we average over larger
distance scales, this contribution will always cancel out, so it can
be ignored in the square barrier limit.

\subsection{The $\sech^2$ potential in $3+1$ dimensions}
\label{sec:sech2}

\begin{figure}
\begin{center}
\leavevmode\epsfbox{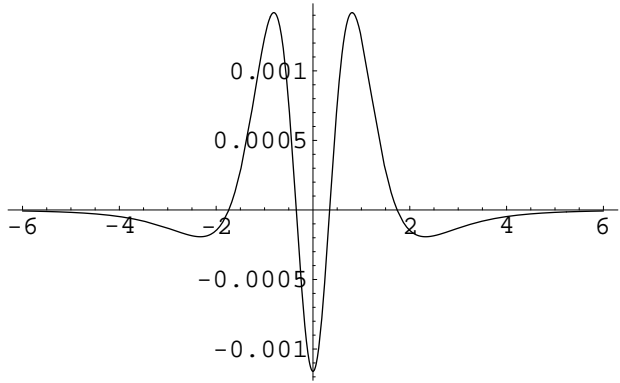}
\end{center}
\caption{Quantum energy density due to the potential of Eq.\
(\ref{eqn:potential}), for $c = 1/a$, $\mu = 0$,  and $M^2={M'}^2=-1/a^2$ in
units where $a=1$.}
\label{fig:densityplot}
\end{figure}

Finally, we consider the potential analyzed in $2+1$ dimensions in
\cite{letter},
\be
V(x) = c^2\sech^2(x/a) \,,
\label{eqn:potential}
\ee
which arises frequently in soliton models.  It is exactly solvable in
terms of associated Legendre functions.  For $c^2 a^2 = -\ell(\ell+1)$ with
integer $\ell$ it becomes reflectionless.  The Green's function at
coincident points is
\be
G(x,x,i\kappa)=
\frac{a}{2}\Gamma (1 + \kappa a + s)\Gamma (\kappa a - s)
{\rm P}^{-\kappa a}_s (\tanh (x/a)) 
{\rm P}^{-\kappa a}_s (-\tanh (x/a))\,.
\label{eqn:Green}
\ee
where ${\rm P}^\mu_\nu (x)$ is the associated Legendre
function as defined in \cite{Bateman:v1} for $-1 < x < 1$, and $s =
(\sqrt{1- 4 c^2 a^2} -1)/2$.  Plugging this into Eq.\
(\ref{eqn:m=1n=2}), we have
\bea
\langle{\cal H}\rangle_{\text{ren}}
&=& -\frac{1}{12\pi^2}
\int_\mu^\infty d\kappa \, \Omega^3 \left[
\frac{a}{2}\Gamma (1 + \kappa a + s)\Gamma (\kappa a - s) \phantom{\frac{1}{1}}
\right. \cr && \times \left.
\left(  2\kappa- \frac{3\kappa}{2\Omega^2} \frac{d^2}{dx^2} \right)
{\rm P}^{-\kappa a}_s (\tanh (x/a))
{\rm P}^{-\kappa a}_s (-\tanh (x/a))
- 1 + \frac{c^2\sech^2(x/a)}{2\kappa^2}
\right. \cr &&- \left.
\frac{c^4\sech^4(x/a)(12 \kappa^2 - M^2)}{2\kappa^2(4\kappa^2 - M^2)^2}
+ \frac{c^2 \sech^2 (x/a) (3 \sech^2 (x/a)-2)} 
{2a^2 (\kappa^2-{M'}^2) \Omega^2} \,,
\right]
\eea
which can then be computed numerically.  Figure\ \ref{fig:densityplot}
gives this energy density as a function of $x$ for particular values of
the parameters.

\section{Conclusions}

We have seen how to address the question of generation of negative energies
through quantum fluctuations in the robust language of quantum field
theory, where ambiguities associated with idealized boundary conditions are
absent.  This approach implements standard renormalization procedures and
is applicable to generic background potentials that are spherically
symmetric in some dimensions and independent of the rest.  Such potentials
typically arise, for example, from topological defects or other extended
objects.  By using dimensional regularization, we have implemented a
precise renormalization scheme, using only local subtractions for both the
first- and second-order diagrams.  We expect that this general formalism,
together with fermion scattering theory in fractional dimensions developed in
\cite{method2}, will allow these results to be extended to fermions and
gauge fields.

In the case of the square barrier, we have recovered the negative energy
associated with perfect reflection at large distances from the barrier, and
we have seen that the divergent negative energy outside the barrier is
canceled by positive energy immediately inside.  In a realistic
example in which one includes the energy associated with the
background potential, such cancellations might lead the averaged null
energy condition to be obeyed even though the weak energy condition is
violated \cite{letter}.  Finally, we have calculated the energy
density for a smooth background representing a domain wall in $3+1$
dimensions.

\section{Note Added in Proof}

Ref.\ \cite{Rebhan} has calculated the surface tension for a bosonic
$\phi^4$ kink domain wall (and also its supersymmetric generalization)
using an on-shell renormalization scheme, in space dimension one through
four.  In the language of the present paper, this calculation corresponds
to the case of $a=2/\mu$ and $c^2 = -3\mu^2/2$ in the potential of Section
\ref{sec:sech2}, renormalized with $M = \mu$, setting $m=1$ and $n$ to zero through
three.  The surface tension is obtained by then integrating this result
over the one nontrivial dimension.  (The choice of $M'$ does not affect
this calculation because the total derivative term integrates to zero.) 
Using the formulae in the present paper to carry out this calculation, we
obtain results in agreement with the bosonic calculations in Ref.\
\cite{Rebhan}.

\section{Acknowledgments}

We would like to thank J. J. Blanco-Pillado, Ruben Cordero and Delia Perlov
for assistance and Larry Ford, Bob Jaffe, Vishesh Khemani, Markus Quandt,
Tom Roman, Marco Scandurra, Xavier Siemens, Alexander Vilenkin, Herbert
Weigel and an anonymous referee for helpful conversations and suggestions. 
N.\ G. was\ supported in part by the U.S.~Department of Energy (D.O.E.)
under cooperative research agreement \#DE-FG03-91ER40662.  K.\ D.\ O. is
supported in part by the National Science Foundation.

\appendix
\section{Calculation of $f$}
\label{sec:fderivation}

By comparing the last line of Eq.\ (\ref{eqn:cterm}) with the first
line of Eq.\ (\ref{eqn:ctermrot}), we require
\be
\frac{1}{\Gamma\left(\frac{n+3}{2}\right)}
\int_\mu^\infty \Omega^{n+1} \kappa^{m-1}
\frac{\sqrt{\pi} f(\kappa,M)}{\Gamma\left(\frac{m}{2}\right)}
  \, d\kappa
= \frac{1}{\Gamma\left(\frac{n+m}{2}\right)}
\int_0^\infty \frac{q^{n+m-1}}{\omega (4\omega^2-M^2)} \, dq \,,
\ee
where $\omega =\sqrt{q^2+\mu^2}$ and $\Omega =\sqrt{\kappa^2-\mu^2}$.

Let us change variables on the left from $\kappa$ to $L =\Omega^2
=\kappa^2-\mu^2$ and on the right from $q$ to $L = q^2$ to get
\be
\frac{1}{\Gamma\left(\frac{n+3}{2}\right)}
\int_0^\infty L^{(n+1)/2} \kappa^{m-2}
\frac{\sqrt{\pi}f(\kappa,M)}{\Gamma\left(\frac{m}{2}\right)} \, dL
= \frac{1}{\Gamma\left(\frac{n+m}{2}\right)}
\int_0^\infty \frac{L^{(n+m-2)/2}}{\omega (4\omega^2-M^2)} \, dL
\ee
with $\kappa =\sqrt{L+\mu^2}$ on the left and $\omega =\sqrt{L+\mu^2}$ on the
right.  We can write
\be\label{eqn:goal}
\frac{1}{\Gamma(1+\alpha)}\int_0^\infty dL\, L^\alpha h (L+\mu^2)
= \frac{1}{\Gamma(1+\beta)}\int_0^\infty dL\, L^\beta j (L+\mu^2)
\ee
with
\bea
\alpha &=& (n+1)/2\\
\beta & = & (n+m-2)/2\\
h (x) &=& \frac{\sqrt{\pi}x^{m/2-1} f(\sqrt{x}, M)}{\Gamma\left(\frac{m}{2}\right)}\\
j (x) &=& \frac{1}{\sqrt{x} (4x - M^2)} \,.
\eea
Denote the difference in the exponents as $\delta =\alpha -\beta =
(3-m)/2$.  If $\delta$ is a positive integer, the desired relationship is just
integration by parts, and
\be
h (x) = \left(-\frac{d}{dx}\right)^\delta j (x) \,.
\ee
To extend this formula to non-integer $\delta$, we write $j$ in terms
of the hypergeometric function,
\be\label{eqn:jF}
j (x) = \frac{1}{2 \sqrt{\pi}}\left(\frac{4}{4x - M^2}\right)^d
\Gamma(d)\, {}_2F_1\left(\frac{1}{2}, d;\frac{3}{2};
\frac{M^2}{M^2 - 4 x} \right) \,,
\ee
with $d=3/2$.  The operator $(-d/dx)$ just increments $d$ in  Eq.\
(\ref{eqn:jF}), so we conjecture that the same relationship holds for all
$\delta$, and thus that the desired $h$ is given by Eq.\ (\ref{eqn:jF})
with $d = 3/2+\delta = 3-m/2$.  One can check that the conjecture is
correct by explicitly performing the integrals in Eq.\ (\ref{eqn:goal}),
which both give
\be
\frac{1}{2\sqrt{\pi}}\left(\mu^2-\frac{M^2}{4}\right)^{\beta-1/2}
\Gamma\left(\frac{1}{2} -\beta\right)\,
{}_2F_1\left(\frac{1}{2},\frac{1}{2}-\beta; \frac{3}{2};
\frac{M^2}{M^2-4\mu^2} \right) \,.
\ee
Finally we find
\bea
f (\kappa, M) &= & \frac{1}{2\pi}\Gamma\left(\frac{m}{2}\right)
\Gamma\left(3-\frac{m}{2}\right)
\left(\frac{4}{4\kappa^2-M^2}\right)^{3-m/2}\kappa^{2-m}
\, {}_2F_1\left(\frac{1}{2},3-\frac{m}{2};\frac{3}{2};
\frac{M^2}{M^2 - 4\kappa^2}\right)\cr
& = & \frac{2(m-4) (m-2)(2\kappa)^{2-m}}{(4\kappa^2 -M^2)^{3-m/2}
\sin\frac{\pi m}{2}} \,{}_2F_1\left(\frac{1}{2},3-\frac{m}{2};\frac{3}{2};
\frac{M^2}{M^2-4\kappa^2}\right) \,.
\eea

\section{Local sum rules}
\label{sec:localsum}

\subsection{General case}

We used the analytic properties of the Green's function as a
mathematical tool, enabling us to carry out calculations
efficiently on the imaginary axis.  In so doing, we avoided the
apparent singularity in the gamma function coefficient of 
Eq.\ (\ref{eqn:e1sub}) for odd $n$.  Nonetheless, this expression should
be a valid result, finite for $m+n<3$.  As in the case of the total
energy \cite{domainwall}, the quantity in brackets must vanish for
$n=1$ in each partial wave individually.  Furthermore, the combination
of the first two terms in brackets vanishes separately from the total
derivative term.  These cancellations depend on a local analog of the
sum rules for the phase shift given in \cite{sumrules1,sumrules2},
which we demonstrate below.

When similar apparent divergences arise in the calculation of the total
energy, they are canceled according to generalizations of
Levinson's theorem \cite{sumrules1,sumrules2}.  For a system with
spherical symmetry, in each partial wave $\ell$ these sum rules take
the form
\be
\sum_j (-\kappa_{\ell j}^2)^N + \int_0^\infty k^{2N} \frac{d}{dk}
\left(\delta_\ell(k) - \sum_{s=1}^N \delta_\ell^{(s)}(k) \right)
 \, dk = 0 \,,
\label{eqn:sumrule}
\ee
where the bound states have $k_{\ell j} = i \kappa_{\ell j}$,
$\delta_\ell(k)$ is the scattering phase shift, and $\delta_\ell^{(s)}(k)$
is the scattering phase shift computed at order $s$ in the Born
approximation.\footnote{In general, these identities continue to hold
even if one subtracts $N'$ orders in the Born approximation for any
$N'\geq N$.  However, there are some restrictions on such
oversubtractions in the symmetric channel in one dimension
\cite{sumrules2}.}  The $N=0$ case gives Levinson's theorem.  Like
Levinson's theorem, these identities apply to general potentials in
scattering theory and hold in each partial wave $\ell$ individually.
Also like Levinson's theorem, they are modified for the case of the
symmetric channel in one dimension, as discussed in \cite{sumrules2}.

We have a relationship \cite{super1d,MITgroup} between the phase
shift, the change in the density of states, and the norm of the
wavefunction,
\be
\frac{1}{\pi} \frac{d\delta_\ell(k)}{dk} =
\rho_\ell(k) - \rho_\ell^{(0)}(k) =
\frac{2 \pi^{\frac{m}{2}}}{\Gamma\left(\frac{m}{2}\right)}
\frac{1}{\pi} \int_0^\infty dr \, r^{m-1} \left(
|\psi_k^\ell(r)|^2 - |{\psi_k^\ell}^{(0)}(r)|^2\right) \,,
\ee
where the zero superscript indicates a quantity evaluated in the free case.
This equation also holds order by order in the Born approximation.
Using these relations we can rewrite Eq.\ (\ref{eqn:sumrule}) as
\be
\frac{2 \pi^{\frac{m}{2}}}{\Gamma\left(\frac{m}{2}\right)}
\int dr \, r^{m-1} \left(
\sum_j (-\kappa_{\ell j}^2)^N |\psi_j^\ell(r)|^2
+ \frac{1}{\pi} \int_0^\infty k^{2N}
\left(|\psi_k^\ell(r)|^2 - \sum_{s=0}^N |{\psi_k^\ell}^{(s)}(r)|^2
\right) dk \, \right) = 0 \,,
\label{eqn:intsum}
\ee
where ${\psi_k^\ell}^{(s)}(r)$ is the Born approximation to the
wavefunction computed at order $s$ (the free wavefunction is the order zero
term).  The identities we need for the present application are simply the
slightly stronger condition that Eq.\ (\ref{eqn:intsum}) holds for each $r$
individually, rather than just as an integral.  We can exploit the
connection to the Green's function that was used in
\cite{sumrules1,sumrules2} to prove this result as well.

The case of $N=0$ is particularly simple, because we know that
\be
\sum_j |\psi_j^\ell(r)|^2 + \frac{1}{\pi} \int_0^\infty \left(
|\psi_k^\ell(r)|^2 - |{\psi_k^\ell}^{(0)}(r)|^2 \right) \, dk = 0
\label{eqn:completeness}
\ee
by completeness; it is just the difference between the expectation
value of a constant computed in the free and interacting bases.  (After
summing over the spectrum, each term is independent of $r$.) For higher
$N$, we would like to show that
\be
\sum_j (-\kappa_{\ell j}^2)^N |\psi_j^\ell(r)|^2
+ \frac{1}{\pi} \int_0^\infty k^{2N}
\left(|\psi_k^\ell(r)|^2 - \sum_{s=0}^N |{\psi_k^\ell}^{(s)}(r)|^2 
\right) \, dk
\ee
is zero.\footnote{As shown in \cite{SUNYSB}, for reflectionless potentials
in one dimension there is a stronger version of the first local sum rule,
\be
\sum_j |\psi_j(r)|^2 \frac{2\kappa_j}{\kappa_j^2 + k^2}
+ |\psi_k(r)|^2 - |{\psi_k}^{(0)}(r)|^2 = 0 \,,
\ee
which reduces to Eq.\ (\ref{eqn:completeness}) when integrated over $k$. 
It might be possible to find analogous results for the higher
sum rules as well.}  We employ the relationship in Eq.\
(\ref{eqn:normtogreen}) between the norm of the wavefunction and the
Green's function to rewrite this expression as
\be
\sum_j (-\kappa_{\ell j}^2)^N |\psi_j^\ell(r)|^2
+ \frac{1}{\pi}
\int_{-\infty}^\infty k^{2N+1} \Im \left(G_\ell(x, x, k) 
- \sum_{s=0}^N G_\ell^{(s)}(x, x, k) \right) \, dk \,,
\ee
where we have extended the integral to the entire $k$ axis by the symmetry
of the integrand.  To show this expression is zero, we would like to
do the $k$ integral as a contour, closed in the upper half plane.  The
singularities in the full Green's function correspond to bound states,
and will exactly cancel the explicit contribution from the bound
states \cite{Bordag}.  The Born approximation has no singularities
(since it does not see the bound states).  Thus we are left with the
contour at infinity.  However, it does not contribute because we have
subtracted enough Born approximations to ensure that the integrand falls like
$1/|k|^2$ at large $|k|$ \cite{scattering}.

\subsection{The symmetric channel}

In one dimension, we have to consider the symmetric channel, which can
have additional singularities at $k=0$.  Such singularities, for
example, lead to an extra $1/2$ in Levinson's theorem \cite{sumrules2},
relating the phase shift at $k=0$ to the number of bound states.  We have
\be
\delta_S(0) = \pi\left(n_S - \frac{1}{2}\right)
\label{eqn:levs}
\ee
as opposed to the usual
\be
\delta(0) = \pi n \,.
\label{eqn:leva}
\ee
Analogously in our problem, Eq.\ (\ref{eqn:completeness}) must be
modified to
\be
\sum_j |\psi_j^S(x)|^2 - \frac{1}{2L} + \frac{1}{\pi} \int_0^\infty \left(
|\psi_k^S(r)|^2 - |{\psi_k^S}^{(0)}(x)|^2 \right) \, dk = 0 \,,
\label{eqn:symcompleteness}
\ee
where $L$ is the size of the system.  Subtracting $\frac{1}{2L}$ reflects
the contribution from the state $\psi(x)=\hbox{const}$ in the free
spectrum.  This state is ``half-bound'':  While any potential
will have a $k=0$ state in the symmetric channel, in this case the
wavefunction goes to a constant at infinity.  Such states contribute
to the spectrum with half the usual residue for a bound state, as the
name indicates.  (Generically a state with $k=0$ will approach a line
with nonzero slope, in which case no special treatment is necessary.)
If a potential has a half-bound state, making the potential
arbitrarily more attractive introduces a new bound state in the
theory, and making it arbitrarily more repulsive eliminates the
half-bound state.\footnote{A \emph{reflectionless} potential will
always have a half-bound state, because it must have $\delta_S(k) =
\delta_A(k)$ for all $k$.  If this equality is to hold at $k=0$, to
reconcile Eqs.\ (\ref{eqn:levs}) and\ (\ref{eqn:leva}) there must be a
half-bound state, which contributes only a half to the number of bound
states.  The half-bound state in the free case (which is
reflectionless) is just a consequence of this requirement.}  There will
be an analogous contribution to the energy density, so this term will cancel
when we pass from Eq.\ (\ref{eqn:nolevsub}) to
Eq.\ (\ref{eqn:yeslevsub}) and the rest of the derivation of the energy
density is unchanged.

For the other sum rules needed in our problem, however, we always
multiply by enough powers of $k$ to cancel any anomalous effects coming
from states at $k=0$.  We would have to be more careful if we do
additional Born ``oversubtractions,'' in which case we could encounter
additional terms analogous to those found in \cite{sumrules2}.  We can
always avoid these problems as long as each ultraviolet Born subtraction is
preceded by a corresponding infrared Levinson subtraction.  For the
first Born subtraction, the corresponding Levinson subtraction was
done using Eq.\ (\ref{eqn:completeness}).  Higher Levinson subtractions
would use local analog of the higher sum rules in \cite{domainwall}.

\subsection{Local subtraction}

For Casimir calculations it will be convenient to slightly modify the
$N=1$ sum rule.  Our renormalization procedure subtracts not the full
first Born approximation, but rather just a local part of it.
However, this replacement does not affect the sum rule.  For example,
to apply the results of Section\ \ref{sec:onedirectsub} for $m=n=1$, we
write
\bea
\sum_{\chi=+,-} |\psi_k(x)|^2
&=& 1 + \int_x^\infty dy \frac{V(y)}{k} \sin 2k(y-x) + \cdots \cr
&=& 1 + \frac{V(x)}{2k^2} + \int_x^\infty dy \frac{V'(y)}{2k^2}
\cos 2k (y-x) + \cdots \cr
&=& 1 + \frac{V(x)}{2k^2} - \int_x^\infty dy \frac{V''(y)}{4k^3}
\sin 2k (y-x) + \cdots \cr
&=& 1 + \frac{V(x)}{2k^2} - \frac{V''(x)}{8k^4}
- \int_x^\infty dy \frac{V'''(y)}{8k^4} \cos 2k (y-x) + \cdots \cr
&=& 1 + \frac{V(x)}{2k^2} - \frac{V''(x)}{8k^4}
+ \int_x^\infty dy \frac{V''''(y)}{16k^5} \sin 2k (y-x) + \cdots
\label{eqn:asymp1d}
\eea
and subtract only the term directly proportional to $V(x)$, rather
than all terms that are first order in the strength of the potential.
However, the additional terms, proportional to the derivatives of $V(x)$,
do not introduce any singularities in the integral and do not affect the
contour at infinity because they fall like $1/k^4$ or faster. Therefore,
this modification does not affect the proof of the sum rule.  This result
allows us to apply the sum rule to Eq.\ (\ref{eqn:e1sub}).

\subsection{One irrelevant dimension}

With the sum rules in hand, we can now extract a finite result from
Eq.\ (\ref{eqn:e1sub}).  Near $n = 1$ we have
\be
\Gamma\left(-\frac{n+1}{2}\right) \approx \frac{2}{n-1} \hbox{~~~~and~~~~}
a^{n-1} \approx (1+\frac{n-1}{2} \log a^2)
\ee
so that in the $n \to 1$ limit we obtain
\bea
\langle{\cal H}\rangle_{\text{ren}} &=& 
-\frac{1}{8\pi} \sum_{\ell} D^m_\ell \left[
\sum_j (\omega_j^\ell)^2 \log (\omega_j^\ell)^2 |\psi^\ell_j(r)|^2
\right. \cr &&+ \left.
\int_0^\infty \frac{dk}{\pi} \omega^2 \log \omega^2
\left( |\psi^\ell_k(r)|^2 - |\psi^\ell_k{}^{(0)}(r)|^2
\left(1 + (2-m)\frac{V(r)}{2k^2}\right) \right)
\right. \cr &&+ \left.
\frac{1}{2} D^2_r \left(
\sum_j \log (\omega_j^\ell)^2 |\psi^\ell_j(r)|^2 + \int_0^\infty
\frac{dk}{\pi} \log \omega^2 |\psi^\ell_k(r)|^2 \right) \right] \,.
\label{eqn:sumrulen=1}
\eea

The local sum rule ensures that the scale of the logarithm does not affect
the final result.  In addition, the limit $\mu\to 0$ is smooth (except when
$n=0$ and $m=1$, where we have the usual infrared divergences of
one-dimensional field theory).  If we extend the range of integration in
Eq.\ (\ref{eqn:sumrulen=1}) as in Eq.\ (\ref{eqn:contourextend}), and then
close the contour in the upper half plane, the branch cut associated with
$\log\omega^2$ will reproduce Eq.\ (\ref{eqn:greenn=1}).


\begin{thebibliography}{99}

\bibitem{Hawking}
S.\ W.\ Hawking, Phys.\ Rev.\ {\bf D46} (1992) 603.

\bibitem{superluminal}
K.\ D.\ Olum, Phys.\ Rev.\ Lett.\ {\bf 81} (1998) 3567.

\bibitem{HelferLang}
A.\ D.\ Helfer and A.\ S.\ Lang, J.\ Phys.\ {\bf A32} (1999) 1937. 

\bibitem{Sopova}
V.\ Sopova and L.\ H.\ Ford, Phys.\ Rev.\ {\bf D66} (2002) 045026.

\bibitem{FordRoman}
L.\ H.\ Ford and T.\ A.\ Roman, Phys.\ Rev.\ {\bf D51} (1995) 4277.

\bibitem{domainwall}
N.\ Graham, R.\ L.\ Jaffe, M.\ Quandt, and H.\ Weigel,
Phys. Rev. Lett.  {\bf 87} (2001) 131601.

\bibitem{method}
E.\ Farhi, N.\ Graham, P.\ Haagensen and R.\ L.\ Jaffe, Phys.\ Lett.\
{\bf B427} (1998) 334; E.\ Farhi, N.\ Graham, R.\ L.\ Jaffe, and H.\ Weigel,
Nucl.\ Phys.\ {\bf B585} (2000) 443.

\bibitem{method2}
E.\ Farhi, N.\ Graham, R.\ L.\ Jaffe, and H.\ Weigel,
Nucl. Phys. {\bf B595} (2001) 536.

\bibitem{SUNYSB}
A.\ S.\ Goldhaber, A.\ Litvintsev and P.\ van Nieuwenhuizen,
Phys.\ Rev.\ {\bf D67} (2003) 105021.

\bibitem{letter}
K.\ D.\ Olum and N.\ Graham,
Phys.\ Lett.\ {\bf B554} (2003) 175.

\bibitem{MITgroup}
N.\ Graham, R.L.\ Jaffe, V.\ Khemani, M.\ Quandt, M.\ Scandurra, and
H.\ Weigel, Nucl.\ Phys.\ {\bf B645} (2002) 49.

\bibitem{super1d}
N.\ Graham and R.L.\ Jaffe, Nucl. Phys.\ {\bf B544} (1999) 432;

\bibitem{Collins}
J.\ C.\ Collins, Phys. Rev.\ {\bf D14} (1976) 1965.

\bibitem{Barcelo:2000zf}
C.\ Barcelo and M.\ Visser, Class. Quant. Grav. {\bf 17} (2000) 3843.

\bibitem{LL}
L.\ D.\ Landau and E.\ M.\ Lifshitz, {\sl Quantum mechanics:
Non-relativistic Theory} (Pergamon Press, Elmsford, NY, 1991), p.\ 214.

\bibitem{Bordag}
M.\ Bordag, J.\ Phys.\ {\bf A28} (1995) 755.

\bibitem{Barton}
G. Barton, J.\ Phys.\ A: Math.\ Gen.\ {\bf 18} (1985) 479.

\bibitem{Kennedy:ar}
G.\ Kennedy, R.\ Critchley and J.\ S.\ Dowker,
Ann. Phys.\  {\bf 125}, 346 (1980).

\bibitem{Ford:1998he}
L.\ H.\ Ford and N.\ F.\ Svaiter,
Phys.\ Rev.\ {\bf D58}, 065007 (1998).

\bibitem{Coleman}
S.\ Coleman and E.\ Weinberg, Phys.\ Rev.\ {\bf D7} (1973) 1888.

\bibitem{Bateman:v1}
Bateman manuscript project, {\sl Higher transcendental functions}
(McGraw-Hill, New York, 1953), volume 1.

\bibitem{Rebhan}
A.\ Rebhan, P.\ van\ Nieuwenhuizen,
R.\ Wimmer, New\ J.\ Phys.\ {\bf 4} (2002) 31.

\bibitem{sumrules1}
R.\ D.\ Puff, Phys. Rev.\ {\bf A11} (1975) 154.

\bibitem{sumrules2}
N.\ Graham, R.\ L.\ Jaffe, M.\ Quandt, H.\ Weigel, 
Ann.\ Phys.\ {\bf 293} (2001) 240.

\bibitem{scattering} R.\ Newton, {\sl Scattering Theory of Waves and
Particles}  (McGraw-Hill, New York, 1966), p.\ 239.

\end{thebibliography}
\end{document}